\documentclass[aps,pre,twocolumn,superscriptaddress,floatfix]{revtex4}
 \usepackage{graphicx,bm,amssymb,amsmath,dcolumn,hyperref}
 \usepackage{multirow}
 \usepackage{enumerate}
 \usepackage{color}
 \usepackage{epstopdf}
 \usepackage{bbm}
 \usepackage{graphicx}
 \usepackage{hyperref}
 \usepackage{threeparttable}
 \usepackage{amsthm}
 \usepackage{mathtools}
 
 \usepackage{mathrsfs}
 \usepackage{subfigure}

\begin{document}
\newcommand{\upcite}[1]{\textsuperscript{\textsuperscript{\cite{#1}}}}
\newcommand{\be}{\begin{equation}}
\newcommand{\ee}{\end{equation}}
\newcommand{\half}{\frac{1}{2}}
\newcommand{\ith}{^{(i)}}
\newcommand{\im}{^{(i-1)}}
\newcommand{\gae}
{\,\hbox{\lower0.5ex\hbox{$\sim$}\llap{\raise0.5ex\hbox{$>$}}}\,}
\newcommand{\lae}
{\,\hbox{\lower0.5ex\hbox{$\sim$}\llap{\raise0.5ex\hbox{$<$}}}\,}

\definecolor{blue}{rgb}{0,0,1}
\definecolor{red}{rgb}{1,0,0}
\definecolor{green}{rgb}{0,1,0}
\newcommand{\blue}[1]{\textcolor{blue}{#1}}
\newcommand{\red}[1]{\textcolor{red}{#1}}
\newcommand{\green}[1]{\textcolor{green}{#1}}

\newcommand{\scrA}{{\mathcal A}}
\newcommand{\scrE}{{\mathcal E}}
\newcommand{\scrF}{{\mathcal F}}
\newcommand{\scrL}{{\mathcal L}}
\newcommand{\scrM}{{\mathcal M}}
\newcommand{\scrN}{{\mathcal N}}
\newcommand{\scrS}{{\mathcal S}}
\newcommand{\scrs}{{\mathcal s}}
\newcommand{\scrP}{{\mathcal P}}
\newcommand{\scrO}{{\mathcal O}}
\newcommand{\scrR}{{\mathcal R}}
\newcommand{\scrC}{{\mathcal C}}
\newcommand{\scrV}{{\mathcal V}}
\newcommand{\scrD}{{\mathcal D}}
\newcommand{\df}{d_{\rm f}}
\newcommand{\dB}{d_{\rm B}}
\newcommand{\dmin}{d_{\rm min}}
\newcommand{\yt}{y_{\rm t}}
\newcommand{\yh}{y_{\rm h}}
\newcommand{\yr}{y_{\rm r}}
\newcommand{\myeq}{ \! = \! }
\newcommand{\rhojunction}{\rho_{\rm j}}
\newcommand{\rhojunctionLim}{\rho_{{\rm j},0}}
\newcommand{\rhobranch}{\rho_{\rm b}}
\newcommand{\rhobranchLim}{\rho_{{\rm b},0}}
\newcommand{\rhononbridge}{\rho_{\rm n}}
\newcommand{\rhononbridgeLim}{\rho_{{\rm n},0}}
\newcommand{\percolationCluster}{C}
\newcommand{\leafFreeCluster}{C_{\rm \ell f}}
\newcommand{\bridgeFreeCluster}{C_{\rm bf}}
\newcommand{\bigO}{\mathcal{O}}

\title{Critical exponents and universal excess cluster number \\ of percolation in four and five dimensions}
\date{\today}

\author{Zhongjin Zhang}
\affiliation{School of Physics and Materials Science, Anhui University, Hefei, Anhui 230601, China}
\affiliation{Hefei National Laboratory for Physical Sciences at the Microscale and
Department of Modern Physics, University of Science and Technology of China,
Hefei, Anhui 230026, China}
\author{Pengcheng Hou}
\affiliation{Hefei National Laboratory for Physical Sciences at the Microscale and
Department of Modern Physics, University of Science and Technology of China,
Hefei, Anhui 230026, China}
\author{Sheng Fang}
\affiliation{Hefei National Laboratory for Physical Sciences at the Microscale and
Department of Modern Physics, University of Science and Technology of China,
Hefei, Anhui 230026, China}
\author{Hao Hu}
\email{huhao@ahu.edu.cn}
\affiliation{School of Physics and Materials Science, Anhui University, Hefei, Anhui 230601, China}
\author{Youjin Deng}
\email{yjdeng@ustc.edu.cn}
\affiliation{Hefei National Laboratory for Physical Sciences at the Microscale and
Department of Modern Physics, University of Science and Technology of China,
Hefei, Anhui 230026, China}
\affiliation{CAS Center for Excellence and Synergetic Innovation Center in Quantum Information and Quantum Physics, University of Science and Technology of China, Hefei, Anhui 230026, China}

\begin{abstract} 

	We study critical bond percolation on periodic four-dimensional (4D) and five-dimensional (5D) hypercubes 
	by Monte Carlo simulations. By classifying the occupied bonds into branches, junctions and non-bridges, 
	we construct the whole, the leaf-free and the bridge-free clusters using the breadth-first-search algorithm.
	From the geometric properties of these clusters, we determine a set of four critical exponents, including 
	the thermal  exponent $\yt \! \equiv \!1/\nu$, the fractal dimension $\df$, the backbone exponent $\dB$ 
	and the shortest-path exponent $\dmin$. We also obtain an estimate of the excess cluster number $b$ which is 
	a universal quantity related to the finite-size scaling of the total number of clusters.
	The results are $\yt \myeq 1.461(5)$, $\df \myeq 3.044 \, 6(7)$, $\dB \myeq 1.984\,4(11)$, 
	$\dmin \myeq 1.604 \, 2(5)$, $b \myeq 0.62(1)$ for 4D; and $\yt \myeq  1.743(10)$, $\df \myeq  3.526\,0(14)$, 
	$\dB \myeq  2.022\,6(27)$, $\dmin \myeq 1.813\, 7(16)$, $b = 0.62(2)$ for 5D.
	The values of the critical exponents are compatible with or improving over the existing estimates, 
	and those of the excess cluster number $b$ have not been reported before. 
	Together with the existing values in other spatial dimensions $d$, the $d$-dependent behavior of the critical exponents 
	is obtained, and a local maximum of $\dB$ is observed near $d \approx 5$. 
	It is suggested that, as expected, critical percolation clusters become more and more dendritic as $d$ increases.
\end{abstract}

%

\maketitle

\section{Introduction} 

Percolation~\cite{StaufferAharony1994}, as a paradigmatic model for studying phase transitions and critical phenomena, can describe diverse phenomena in various fields such as fluids in porous media, gelation process and epidemiology. Considering bond percolation on a lattice, each edge is occupied independently by a bond with probability $p$ and clusters are constructed as sets of sites that 
are connected via occupied bonds. The clusters become larger as $p$ increases, and at the critical point $p_c$, an infinite cluster spanning over the whole lattice will emerge. Near $p_c$, the clusters exhibit rich behaviors, which are characterized by universal critical exponents~\cite{StaufferAharony1994} and other universal quantities, 
including wrapping probabilities~\cite{Hu15}, dimensionless ratios~\cite{Binder81,Selke07}, and the excess cluster number $b$ related to the finite-size scaling of the total number of clusters~\cite{ziff1997}. 

Values of these universal quantities depend on the spatial dimension $d$. 
For percolation in two dimensions (2D), great advances have been made via 
the renormalization group~\cite{harris1975renormalization,Gracey15, PhysRevE.59.R6239,Janssen2000}, 
the Coulomb-gas method~\cite{PhaseTransitionandCriticalPhenomena2}, the conformal field theory~\cite{PhaseTransitionandCriticalPhenomena} and the stochastic Loewner evolution~\cite{Schramm01}, etc. Many exact results are now available, including the thermal exponent $\yt  \! \equiv \! 1/\nu \myeq 3/4$, which is identical to the red-bond exponent $d_{\rm red}$ for percolation~\cite{Stanley77, Xu2014a}, 
the magnetic exponent $\yh$ which is equivalent to the fractal dimension $\df \myeq 91/48$, the excess cluster number $b \myeq 0.883\,576\,308...$~\cite{Kleban1998}, as well as 
various universal wrapping probabilities~\cite{Hu15}. But there are still critical exponents, such as the backbone exponent $\dB$ and the shortest-path exponent $\dmin$, whose values are solely given numerically~\cite{Xu2014a,PhysRevE.86.061101}.
For dimensions at or above the upper critical dimension $d\ge d_{\rm u} \myeq 6$, mean-field theory~\cite{aharony1984scaling} 
leads to $\yh \myeq \df \myeq 2d/3$, $\yt \myeq \dmin \myeq \dB \myeq d/3$ for critical percolation on periodic hypercubes.
However, for percolation in $2<d<6$ there has been no exact solution for these universal quantities.
Instead one uses approximate or numerical methods, e.g. the $\epsilon$-expansion method~\cite{harris1975renormalization,Gracey15, PhysRevE.59.R6239,Janssen2000}, 
the conformal bootstrap method~\cite{bootstrap18}, and the Monte Carlo (MC) method. 
Using MC simulations, high-precision results for universal quantities have been obtained for 3D~\cite{Xu2014}.  
While for 4D and 5D, existing results of the critical exponents are not as precise as those for 3D, and the excess number $b$ has not been estimated.

In this paper, we determine with high-precision a set of four critical exponents(i.e. $\yt$, $\df$, $\dB$, and $\dmin$) 
and the universal excess cluster number $b$,
for percolation on 4D and 5D hypercubes, using MC simulations and finite-size scaling analysis. 
Our results for $y_{\rm t}$ and $d_{\rm f}$ are compatible with the recent
estimates~\cite{Koza16,Mertens18}, and those for $d_{\rm B}$ and $d_{\rm min}$ 
significantly improve over the existing ones~\cite{Mouk98,Paul01}.
Together with previous results in other spatial dimensions, we obtain
the $d$-dependent behaviors of the critical exponents. It is interesting to see that 
 $d_{\rm B}$ is not a monotonic function of $d$, 
and it has a local maximum near $d \approx 5$.
As $d$ increases, the values of $\yt$, $\dmin$ and $\dB$ approach to each other,
while the differences between them and the $\df$ value get larger, 
suggesting that the critical percolation clusters become more dendritic.

The rest of the paper is organized as follows. 
Section~\ref{simulation-data-analysis} describes the simulation and fitting ansatz. 
Section~\ref{sec:results} presents detailed numerical results. 
In addition to the four critical exponents and the excess cluster number $b$,
results are also presented for two dimensionless ratios based on 
cluster-size distributions and for densities of three different types of bonds.
The $d$-dependent behavior of the critical exponents is 
illustrated in Sec.~\ref{main-results}, and a brief discussion is given in 
Sec.~\ref{Sec:dis}.

\section{Simulation and fitting ansatz}  
\label{simulation-data-analysis}

We consider critical bond percolation with periodic boundary conditions on 
the 4D and 5D $L^d$ hypercubic lattices, where $L$ is the linear size of the system.  
The percolation thresholds are taken as $p_c (4 {\rm D}) \myeq 0.160\, 131\, 22$~\cite{Mertens18,XunZiff19} 
and $p_c (5 {\rm D}) = 0.118\, 171\, 45$~\cite{Mertens18}. 
In 4D, the simulated system sizes are  $L \myeq $ 4,6,8,10,12,14,16,20,24,32,40,48,64, 96,128,176,192,
and the numbers of independent samples are more than $10^6$ for each $L \le 64$  
and  at least $10^5$ for each $L > 64$. 
In 5D, more than $10^6$  samples are taken for each of $L \myeq 4,6,8,10,12,14,16,20$,
and at least $10^5$  samples for each of $L \myeq 24,32,40,48,60$.

In the simulation, we sample the following observables: 
\begin{itemize}
	\item The cluster-number density $n_{c}$. At $p_c$ it scales as 
		\begin{equation}
			n_c \myeq n_{c,0} + b/V + \ldots, 
			\label{Eq:nc}
		\end{equation}
		where $V \myeq L^{d}$ is the volume of the $d$-dimensional hypercube, $b \myeq \lim_{V \rightarrow \infty} V \left( n_c-n_{c,0} \right)$ is the excess number of clusters over the bulk value. Given the spatial dimension, the quantity $b$ is universal and depends only on the system shape and boundary conditions~\cite{ziff1997,ziff1999}. 

	\item The mean size of the largest cluster $C_1 \myeq \left<\mathcal C_1\right>$. It scales as $C_1 \sim L^{\df}$ at $p_c$.

	\item The mean cluster-size moments $\left<\mathcal S_2\right>$ and $\left<\mathcal S_4\right>$. The $m$th-order moment $\mathcal S_m$ is defined as $\mathcal S_m \myeq \sum_i {\mathcal C}_i^m$, where the summation is over all clusters and ${\mathcal C}_i$ denotes the size of a cluster.

	\item The reduced susceptibility $\chi' \myeq \left<\mathcal S_2-\mathcal C_1^2\right> / V$. It scales as $\chi' \sim L^{2\df-d}$ at $p_c$.

	\item Two dimensionless ratios
		\begin{equation}
			Q_1 \myeq \frac{\left<\mathcal C_1^2\right>}{\left<\mathcal C_1\right>^2} 
			\hspace{3mm} \mbox{and} \hspace{3mm} 
			Q_s \myeq \frac{\left<3\mathcal S_2^2-2\mathcal S_4\right>}{\left<\mathcal S_2\right>^2} \;.
		\end{equation}
		They reflect properties of the size distribution of clusters, and at criticality scale as $Q_{i,0}+aL^{y_i}+\ldots$, where $y_i<0$ is an irrelevant correction exponent, 
		and $Q_{i,0}$ represents the infinite-size value, 
		depending on the system shape and the boundary conditions~\cite{Binder81,Selke07}.

	\item The number of occupied bonds $\scrN_{\rm b}$ and the covariances of the dimensionless ratios and $\scrN_{\rm b}$~\cite{deng2003}: 
$$g_{pQ_1} \myeq \frac{2\left<\mathcal C_1\scrN_{\rm b}\right>}{\left<\mathcal C_1\right>}-\frac{\left<\mathcal C_1^2\scrN_{\rm b}\right>}{\left<\mathcal C_1^2\right>}-\left<\scrN_{\rm b}\right> \, ,$$
		\begin{equation}
g_{pQ_s} \myeq \frac{2\left<\mathcal S_2 \scrN_{\rm b}\right>}{\left<\mathcal S_2\right>}-\frac{\left<(3\mathcal S_2^2-2\mathcal S_4) \scrN_{\rm b}\right>}{\left<3\mathcal S_2^2-2\mathcal S_4\right>}-\left<\scrN_{\rm b}\right> \, .
		\end{equation}
The covariances scale as $\sim L^{\yt}$ at $p_c$. 

	\item The mean shortest-path length $S \myeq \left<\mathcal S\right>$. It scales as $S \sim L^{d_{\rm min}}$~\cite{0305-4470-17-8-007} at $p_c$. 
		Here we have 
		\begin{equation}
		\mathcal{S}:= \max_\scrC\max_{y\in \scrC}d\left(x_\scrC,y\right)\,,
		\end{equation}
		where $\scrC$ denotes different clusters, 
		$x_\scrC$ is the original site from which the cluster $\scrC$ grows,
$d\left(x,y\right)$ is the graph distance between two sites in the same cluster. 
		Thus $ \max_{y\in \scrC} d\left(x_\scrC,y\right)$ is the maximum time steps for growing the cluster $\scrC$ from the original site $x_\scrC$ in the breadth-first-growing procedure, and the other `$\max$' selects 
the maximum of $ \max_{y\in \scrC} d\left(x_\scrC,y\right)$ for all clusters. 
		The quantity $S$ is also known as the chemical distance.

\end{itemize}

To further investigate the geometric properties of the clusters, the occupied bonds are classified into three types: branch, junction and non-bridge~\cite{Xu2014a}. If the deletion of an occupied bond causes the cluster to break into two parts, the bond is a bridge; an occupied bond that is not a bridge is a non-bridge.
A bridge is a junction if neither of the two broken parts after its deletion is a tree, and otherwise it is a branch.
Deleting all branches from whole percolation clusters leads to leaf-free clusters, and deleting all bridges produces bridge-free clusters. Accordingly, we measure the following quantities: 

\begin{itemize}

	\item The mean size of the largest leaf-free and bridge-free cluster $C_{\rm lf}$, $C_{\rm bf}$. They scale as  $C_{\rm lf}\sim L^{\df}$~\cite{Zhou15}, $C_{\rm bf} \sim L^{d_{\rm B}}$~\cite{Xu2014a} at $p_c$. 

	\item  The densities of branches, junctions and non-bridges $\rho_{\rm b},\rho_{\rm  j},\rho_{\rm n}$. They scale as $\rho + a L^{\yt-d} + \ldots$ at $p_c$. 

\end{itemize}

It is noted that, different from bridge-free clusters, the backbone of the incipient infinite cluster is traditionally defined as the subset of the cluster carrying the current when a voltage difference is applied between two sides of the system or between two sites far apart~\cite{HerrmannHongStanley1984, Grassberger1999}. However, both definitions of the backbone are essentially equivalent, since they both scale as $\sim L^{\dB}$. 

We perform least-square fits of the MC data for the quantities by the finite-size scaling ansatz
\begin{equation}
\mathcal O \myeq c_0+L^{d_{\mathcal O}}\left(a_0+a_1L^{y_1} + \ldots \right)\,\,,
\end{equation}
where $d_{\mathcal{O}}$ is a critical exponent, $a_0$ is a constant, $a_1L^{y_1}$ is the leading correction term with amplitude 
$a_1$ and exponent $y_1<0$, and $c_0$ is the background term. 
For 4D, a value of $y_1= -\omega = -\Omega \df = -1.22(9)$~\cite{XunZiff19} is recently given by measuring the cluster size distributions in MC simulations. And for 5D, Gracey gives the series expansion result of $y_1=-\omega \simeq -0.72$~\cite{Gracey15}. This work assumes that $y_1$ may be different for different quantities, and does not aim to estimate accurate values of it. Fits are first performed with $y_1$ being a free fitting parameter, then with $y_1$ being fixed at values around the first set of results. If $a_1$ is found to be consistent with zero, fits are also performed without the correction term. And in cases the leading correction term is not adequate to describe the finite-size data, fits are performed by including the subleading correction term $a_2 L^{y_2}$, where $a_2$ is the amplitude and $y_2$ is the subleading correction exponent.

As a precaution against other correction-to-scaling terms which are not included in the fitting ansatz, we impose
a lower cutoff $L\ge L_{{\rm m}}$ on the data points admitted in the fits, and systematically study the effect on the $\chi^2$ value when increasing $L_{{\rm m}}$. Generally, we prefer fits for any given ansatz corresponding to the smallest $L_{{\rm m}}$ for which the goodness of fit is reasonable, and for which subsequent increase in $L_{{\rm m}}$ does not cause the $\chi^2$ value to drop vastly by more than one unit per degree of freedom. 
In practice, by `reasonable' we mean that $\chi^2/{\rm DF}\lesssim 1$, where `DF' is the number of degrees of freedom.

The error of our estimates consists of both the statistical and the systematic error. 
The former refers to the error that directly propagates to the fit results from the measured observables, 
and the latter mainly comes from the truncation of fitting ansatz, particularly finite-size corrections.
We perform fits with different $L_{{\rm m}}$ and $y_{{\rm 1}}$, 
and evaluate the difference of the fitting results.
For each quantity, the reliability of the final result is checked by plotting the quoted value as well as 
those with 3-sigma deviations from it.

\section{Numerical results}   
\label{sec:results}
In this section, we present the results for the four critical exponents,
the excess cluster number, the two dimensionless ratios and the bond densities for branches, junctions and non-bridges.

\subsection{Thermal exponent}
\label{sec:the-exp}

To estimate the thermal exponent $y_{\rm t}$, the data of $g_{pQ_s}$ 
are fitted by the ansatz $g_{pQ_s} \myeq L^{y_{\rm t}}(a_0 + a_1 L^{y_1})$. 
Fits are performed with the correction exponent $y_1$ as a free fitting parameter,
and also with $y_1$ fixed at different values.
The correction amplitude $a_1$ is found to be consistent with zero within error bars when $L_{{\rm m}}\ge 10$ for 4D and $L_{{\rm m}}\ge 8$ for 5D, 
suggesting that for large $L_{{\rm m}}$, finite-size corrections for $g_{pQ_s}$ are not significant. 
Thus fits are also made without the correction term, i.e. $a_1 \myeq 0$. 
The results are listed in Table~\ref{Tab:gpqs}, from which we obtain 
$y_{\rm t} \myeq 1.461(5)$ for 4D and $1.743(10)$ for 5D.

\begin{figure}
	\centering
	\includegraphics[scale=1.0]{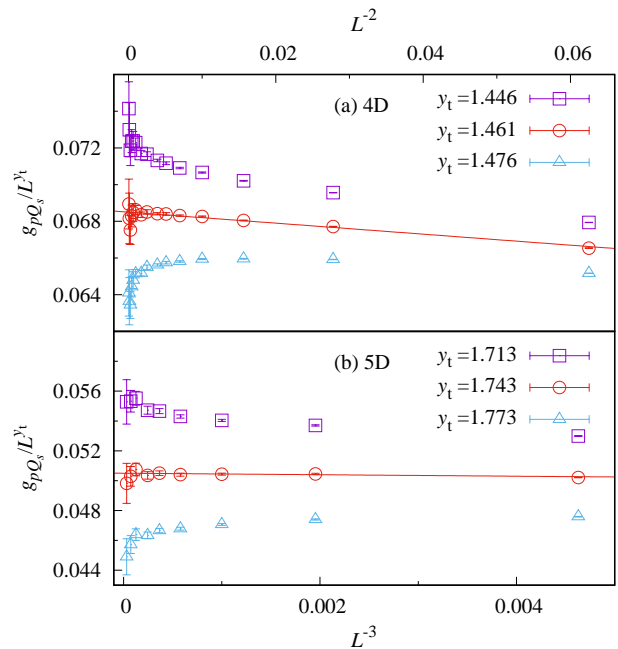}
	\caption{Plots of $g_{pQ_s}/L^{\yt}$ versus $L^{-2}$ or $L^{-3}$ for critical 4D (a) and 5D (b) percolation respectively, illustrating our estimate $\yt(\rm {4D})=1.461(5)$ and $\yt(\rm {5D})=1.743(10)$.
	The reliability of the results is clearly demonstrated by the upward or downward bending 
	for values $\yt \pm 3 \Delta$, with $\Delta$ the error bar.
	The straight lines are obtained from the fits.}
	\label{Fig:gpqs}
\end{figure}

To show the  reliability of the results, we plot $g_{pQ_s}/L^{\yt}$ versus $L^{y_1}$ in Fig.~\ref{Fig:gpqs}, 
where $y_{\rm t}$ is chosen to be the above estimates and the values away from them
by  three error bars ($3\Delta$). Since $g_{pQ_s}\sim L^{y_{\rm t}}$ as $L\rightarrow \infty$, the obvious 
upward (downward) bending as $L$ increases when using the central value minus (plus) three error bars illustrates that the true value of $y_{\rm t}$ falls in the interval  $\left[y_{\rm t}-3\Delta,y_{\rm t}+3\Delta\right]$. 

Fits are also performed for the covariance $g_{pQ_1}$, whose results are also listed in Table~\ref{Tab:gpqs}. Estimates of $\yt$ from  $g_{pQ_1}$ are consistent with those from $g_{pQ_s}$, but with a lower precision.

\begin{table}[htb]
	\centering\small
	\caption{Fitting results for covariances $g_{pQ_1}$ and $g_{pQ_s}$.}
	\label{Tab:gpqs}
	\setlength{\tabcolsep}{0.5mm}{
		\begin{tabular}{|l|l|l|l|c|c|}
			\hline
			$\mathcal O$&$d$	& $\yt$ 	& $a_0$ 	& $y_1$  & $L_{\rm m}/{\rm DF}/\chi^2$\\
			\hline
			&&1.463(1)		&0.068\,1(3)	&-2.6(2)		&4/11/11\\
			&&1.461(4)		&0.068\,6(9)	&-1.8(7)		&6/10/10\\
			&&1.461\,2(16)		&0.068\,5(3)	&-2		&6/13/8 \\
			&4&1.459\,4(25)		&0.068\,9(5)	&-2		&8/12/7\\
			&&1.464\,5(12)		&0.067\,7(2)	&/		&10/12/8\\
			$g_{pQ_s}	$	&&1.463\,9(17)		&0.067\,8(3) &/		&12/11/7\\
			\cline{2-6}
			&&1.737(5)	&0.051\,3(7)	&-3.1(6)	&4/9/12\\
			&&1.738(3)	&0.051\,1(4)	&-4	&6/9/11\\
			&&1.743(6)	&0.050\,4(8)	&-4	&8/8/11\\
			&5&1.736(2)	&0.051\,4(2)	&-3	&4/10/12\\
			&&1.736(3)	&0.051\,4(4)	&-3	&6/9/12\\
			&&	1.742(2)	&0.050\,5(4)	&/	&8/9/5\\
			&&1.743(6)	&0.050\,4(7)	&/	&10/8/5\\
			\hline
			&&1.462(5)		&0.017\,1(3)	&-2.1(2)	&4/13/12\\
			&&1.463(12)		&0.017\,7(8)	&-1.7(5)		&6/12/11\\
			&4&1.462(1)		&0.068\,4(2)	&-2		&6/13/11\\
			$g_{pQ_1}$	&&1.460(2)		&0.068\,8(4)	&-2		&8/12/10\\
			\cline{2-6}
			&&1.71(4)		&0.010\,5(12)	&-2.2(6)		&4/9/5\\
			&&1.70(2)		&0.010\,8(6)	&-2			&6/9/6\\
			&5&1.74(5)		&0.009\,7(13)	&-2			&8/8/5\\
			&&1.74(3)		&0.009\,7(9)	&/			&12/7/4\\
			&&1.74(5)		&0.009\,7(14)	&/			&14/6/4\\
			\hline
	\end{tabular}}
\end{table}

\subsection{Fractal dimension}
To estimate the fractal dimension $\df$,  the data of $C_1$ and $C_{\rm lf}$ 
are fitted to the ansatz $L^{\df}(a_0+a_1L^{y_1})$. 

For $C_1$ of the 4D model, 
stable fitting results cannot be obtained for $y_1$ when setting it as a free fitting parameter. 
While fixing $y_1 \myeq -2$ leads to reasonable fitting results for $L_{{\rm m}}\ge 20$, 
which have $a_1$ being consistent with zero within error bars for $L_{{\rm m}} > 24$. 
If fixing $a_1 \myeq 0$, reasonable fits can be obtained for $L_{{\rm m}} \! > \! 24$.
For $C_1$ of the 5D model, 
setting $y_1$ as a free fitting parameter produces the estimate $y_1 \myeq -2.0\left(5\right)$.
Further fits are performed with fixed $y_1 \myeq -2$.
These fitting results for $C_1$ are listed in Table~\ref{Tab:C1C1lf}, from which we estimate $\df \myeq 3.045\,2(8)$ 
and $3.526\,0(14)$ for the 4D and 5D model, respectively.

Applying similar procedures to $C_{\rm lf}$, we estimate $y_1 \approx -2.71$ for 4D, $-2.4$ for 5D, and obtain $\df \myeq 3.044\,6(7)$ and $3.525\,4(19)$ for 4D and 5D, respectively. These fitting results are also listed in Table~\ref{Tab:C1C1lf}

\begin{figure}
	\centering
	\includegraphics[scale=1.0]{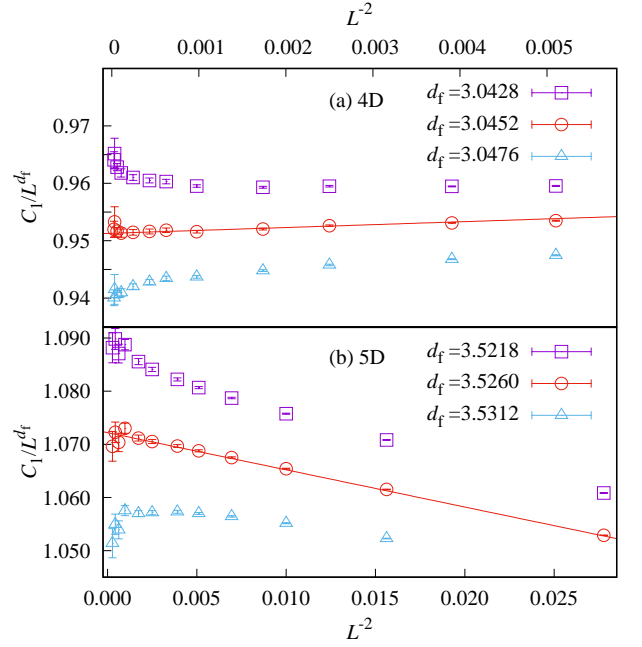}
	\caption{Plots of $C_1/L^{d_{\rm f}}$ versus $L^{-2}$ for critical 4D (a) and 5D (b) percolation, illustrating our estimate $d_{\rm f}(\rm {4D})=3.045\,2(8)$ and $d_{\rm f}(\rm {5D})=3.526\,0(14)$. The straight lines are obtained from the fits.}
	\label{Fig:C1}
\end{figure}

\begin{figure}
	\centering
	\includegraphics[scale=1.0]{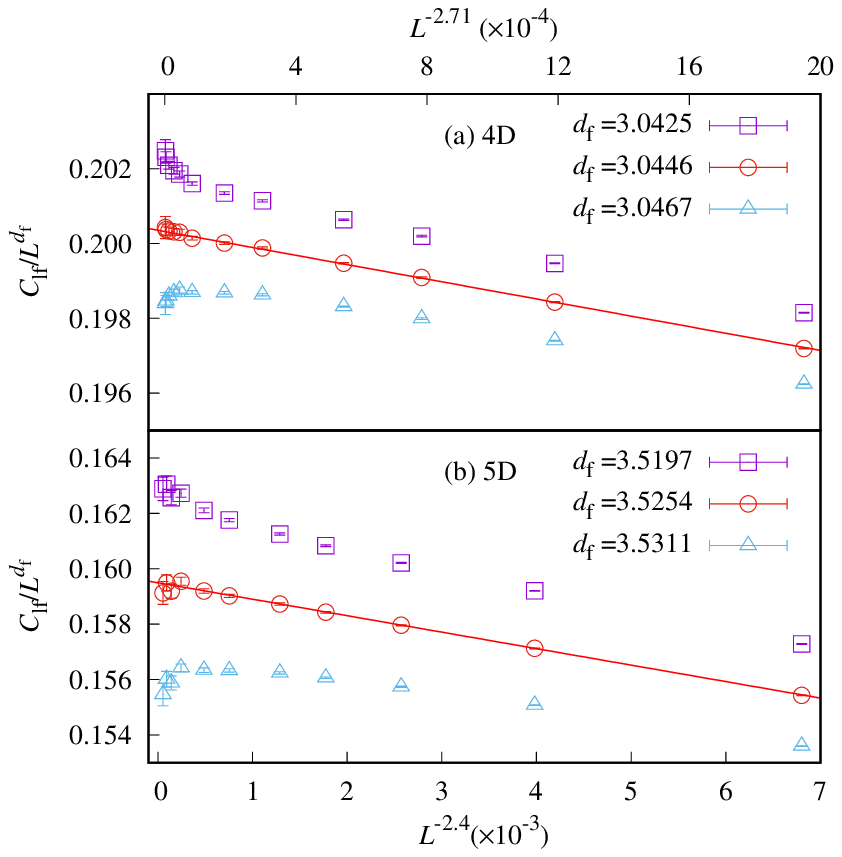}
	\caption{Plots of $C_{\rm {lf}}/L^{d_{\rm f}}$ versus $L^{-2.71}$ or $L^{-2.4}$ for critical 4D (a) and 5D (b) percolation respectively, illustrating our estimate $d_{\rm f}(\rm {4D})=3.044\,6(7)$ and $d_{\rm f}(\rm {5D})=3.525\,4(19)$.The straight lines are obtained from the fits.}
	\label{Fig:Clf}
\end{figure}

We plot $C_1/L^{\df}$ ($C_{\rm lf}/L^{\df}$) versus $L^{y_1}$ in Fig.~\ref{Fig:C1} (Fig.\ref{Fig:Clf}), where $\df$ is chosen to be the central value of the estimates and the central value plus or minus three error bars.  The obvious upward (downward) bending as $L$ increases when using the central value minus (plus) three error bars illustrates the reliability of our estimates for $\df$.

The above estimates of $\df$ from $C_1$ and $C_{\rm lf}$
are consistent with each other within error bars, and they lead  to our final estimate of $\df$ as $3.044\,6(7)$ and $3.526\,0(14)$ for 4D and 5D, respectively.

\begin{table}[htb]
	\centering\small
	\caption{Fitting results for $C_1$ and $C_{\rm lf}$.}
	\label{Tab:C1C1lf}
	\begin{tabular}{|l|l|l|l|c|c|}
		\hline
		$\mathcal O$	&$d$& $\df$ 	& $a_0$ 	&  $y_1$  & $L_{\rm m}/{\rm DF}/\chi^2$\\
		\hline
		&&3.045\,8(3)		&0.949(1)   &-2      &20/7/4\\
		&4&3.045\,6(4)		&0.950(2)   &-2      &24/6/3\\
		&&3.045\,2(4)		&0.952(2)   &/		&32/6/1\\
		$C_{\rm 1}$ &&3.045\,1(5)		&0.952(2) &/		&40/5/1\\
		\cline{2-6}
		&&3.526\,4(10)	&1.071(4)	&-2.0(2)	&6/8/4\\
		&5&3.525\,7(20)	&1.074(8)	&-1.9(4)	&8/7/4\\
		&&	3.525\,7(3)	&1.073(1)	&-2	&6/9/4\\
		&&3.526\,0(5)	&1.072(2)	&-2	&8/8/4\\
		\hline
		&&3.044\,6(2)		&0.200\,3(2)&-2.71(2)	&6/11/7 \\
		&4&3.044\,5(3)		&0.200\,4(3)&-2.68(5)	&8/10/6\\
		$C_{\rm lf}$	&&3.044\,5(2)		&0.200\,4(2)&-2.71		&10/10/5\\
		&&3.045\,1(2)		&0.200\,0(1)& -3		&12/9/7\\
		\cline{2-6}
		&&3.525\,8(15)	&0.159\,3(8)	&-2.5(2)	&8/7/4\\
		&&3.525(3)	&0.159\,8(2)	&-2.3(5)	&10/6/3\\
		&5&	3.525\,2(5)	&0.159\,6(2)	& -2.4	&8/8/4\\
		&&3.525\,5(8)	&0.159\,4(3)	& -2.4	&10/7/3\\
		\hline
	\end{tabular}
\end{table}

The reduced susceptibility scales as $\chi' \myeq L^{2\df-d} (a_0+a_1L^{y_1})$, 
from which one can also estimate $\df$. 
The fitting results are listed in Table~\ref{Tab:chi}, from which we obtain 
estimates $2\df-d \myeq 2.099(9)$ and $2.052(4)$, leading to 
$\df \myeq 3.049(5)$ and $3.526(3)$ for 4D and 5D, respectively. 
These are consistent with results from fitting the data of 
$C_1$ and $C_{\rm lf}$.

	\begin{table}[htb]
	\centering\small
	\caption{Fitting results for $\chi'$.}
	\label{Tab:chi}
	\begin{tabular}{|l|l|l|c|c|}
		\hline
		$d$& $2\df-d$ 	& $a_0$ 	&  $y_1$  	&  $L_{\rm m}/DF/\chi^2$\\
		\hline
		&2.099(6)		&0.232(8)	&-0.72(12)		&16/7/2\\
		&2.099(9)		&0.231(13)	&-0.72(24)		&20/6/2\\
		4&2.099\,0(8)	&0.231\,4(9)	& -0.72			&16/8/2\\
		&2.099\,0(12) 	&0.231(2)	& -0.72			&20/7/2 \\
		\cline{1-5}
		&2.054\,7(11)	&0.536(2)	&-1.70(6)	&6/8/5\\
		5&2.053\,9(19)	&0.538(4)	&-1.7(2)	&8/7/5\\
		&2.053\,3(16)	&0.539(3)	&-1.7		&14/5/4\\
		&2.051\,7(21)	&0.542(4)	&-1.7		&16/4/3\\
		\hline
	\end{tabular}
\end{table}

\subsection{Backbone exponent}

To estimate $\dB$, the data of $C_{\rm bf}$ are fitted to the ansatz $L^{\dB}(a_0+a_1L^{y_1})$. 
For 4D, stable fitting results cannot be obtained when setting $y_1$ as a free fitting parameter. 
When fixing $y_1$, $a_1$ is found to be consistent with zero within error bars 
for large cutoff sizes $L_{\rm m}$. 
Then fits are performed with fixed $a_1 \myeq 0$. For 5D, 
$y_1$ is estimated to be $-1.8\left(2\right)$ when setting it as a free fitting parameter.
Subsequent fits are performed with fixed $y_1 \myeq -1.8$. The above fitting results 
are listed in Table~\ref{Tab:Cbf}, from which we estimate 
$d_{\rm B} \myeq 1.984\,4(11)$ and $2.022\,6(27)$ for 4D and 5D, respectively. 

\begin{figure}
	\centering
	\includegraphics[scale=1.0]{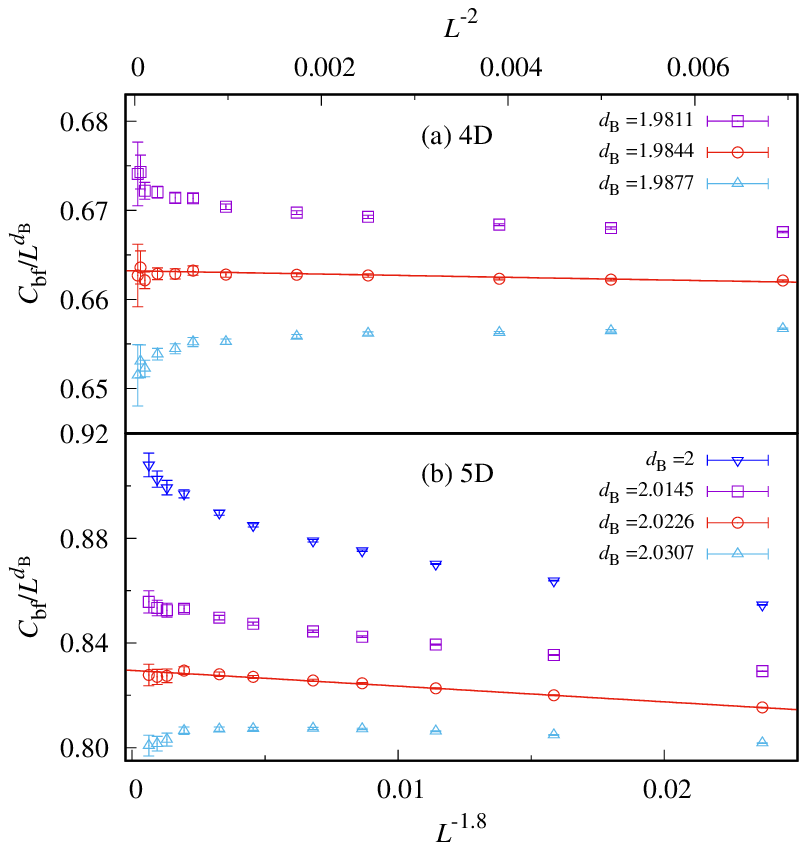}
	\caption{Plots of $C_{\rm {bf}}/L^{d_{\rm B}}$ versus $L^{-2}$ or $L^{-1.8}$ for critical 4D (a) and 5D (b) percolation respectively, illustrating our estimate $d_{\rm B}(\rm {4D})=1.984\,4(11)$ and $d_{\rm B}(\rm {5D})=2.022\,6(27)$.The straight lines are obtained from the fits.}
	\label{Fig:Cbf}
\end{figure}

We plot $C_{\rm bf}/L^{\dB}$ versus $L^{y_1}$ in Fig.~\ref{Fig:Cbf}, where $d_{\rm B}$ 
is chosen to be the central value and the central value plus or minus 
three error bars. The obvious upward (downward) bending as $L$ increases 
when using the central value minus (plus) three error bars illustrates 
the reliability of the estimate for $\dB$.
In Fig.~\ref{Fig:Cbf} we also plot the data assuming $\dB \myeq 2$, 
and see that the curve bend upward as $L$ increases. This tells that 
$\dB$ for 5D is indeed greater than the value $2$ for 6D.

\begin{table}[htb]
	
	\centering\small
	\caption{Fitting results for $C_{\rm bf}$.}
	\label{Tab:Cbf}
	\begin{tabular}{|l|l|l|c|c|}
		\hline
		$d$& $d_{\rm B}$ 	& $a_0$ 	&  $y_1$  	&  $L_{\rm m}/{\rm DF}/\chi^2$\\
		\hline
		&1.984\,7(5)		&0.662(1)	& -4			&14/8/3\\
		&1.984\,1(5)		&0.664(1)	& -2			&14/8/3\\
		4 &1.984\,0(6)		&0.664(2)	& -1			&12/9/5\\
		&1.984\,6(5) 	&0.662(1)		&/			&20/7/2 \\
		&1.984\,5(6)		&0.663(2)	&/			&24/6/2\\
		\cline{1-5}
		&2.022\,7(25)	&0.828(7)	& -1.8(2)	&6/8/3\\
		5&2.023\,1(12)	&0.828(3)	& -1.8	&8/8/2\\
		&2.022\,1(18)	&0.831(5)	& -1.8	&10/7/2\\
		\hline
	\end{tabular}
\end{table}

\subsection{Shortest-path exponent}
To estimate $\dmin$, the data of the quantity $S$
are fitted to the ansatz $S \myeq L^{\dmin} (a_0 + a_1 L^{y_1})$.
Fits are first performed with $y_1$ being a free fitting parameter, 
which leads to $y_1 \myeq -1.84\left(9\right)$ and $-1.65(10)$ for 4D and 5D, respectively.
Subsequent fits are made with $y_1$ fixed at $-1.84$ and $-1.65$ 
for 4D and 5D, respectively. These fitting results are listed in Table~\ref{Tab:S}, 
from which we estimate $d_{{\rm min}} \myeq 1.604\,2(5)$ 
and $1.813\,7(16)$ for 4D and 5D, respectively. 

\begin{figure}
	\centering
	\includegraphics[scale=1.0]{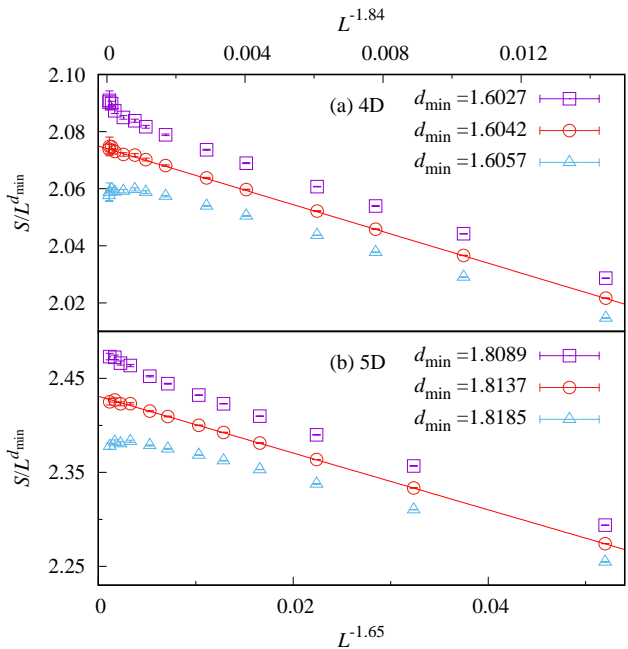}
	\caption{Plots of $S/L^{d_{\rm min}}$ versus $L^{-1.84}$ or $L^{-1.65}$ for critical 4D (a) and 5D (b) percolation respectively, illustrating our estimate $d_{\rm min}(\rm {4D})=1.604\,2(5)$ and $d_{\rm min}(\rm {5D})=1.813\,7(16)$.The straight lines are obtained from the fits.}
	\label{Fig:S}
\end{figure}

We plot $S/L^{\dmin}$ versus $L^{y_1}$ in Fig.~\ref{Fig:S}, where $\dmin$ is chosen to be the central value and the central value plus or minus three error bars. The obvious upward (downward) bending as $L$ increases when using the central value minus (plus) three error bars illustrates the reliability of the estimate for $\dmin$.

\begin{table}[htb]
	\centering\small
	\caption{Fitting results for $S$.}
	\label{Tab:S}
	\begin{tabular}{|l|l|l|c|c|}
		\hline
		$d$	& $d_{\rm min}$ & $a_0$ 	&  $y_1$  	& $L_{\rm m}/{\rm DF}/\chi^2$\\
		\hline
		&1.604\,2(3)		&2.075(2)	& -1.82(4)	&10/10/4 \\
		4&1.604\,3(4)		&2.074(3)	& -1.86(7)	&12/9/4\\
		&1.604\,26(12)	&2.073\,9(8)	& -1.84		&10/11/4\\
		&1.604\,21(15)	&2.074\,3(11)	& -1.84		&12/10/4\\
		\cline{1-5}
		&1.813\,3(8)	&2.435(7)	& -1.63(3)	&6/8/4\\
		5&1.813\,9(14)	&2.429(13)	& -1.67(8)	&8/7/4\\
		&	1.813\,8(2)	&2.430(2)	& -1.65	&6/9/5\\
		&1.813\,6(4)	&2.432(3)	& -1.65	&8/8/4\\
		\hline
	\end{tabular}
\end{table}

\subsection{The excess cluster number}  
\begin{figure}
	\centering
	\includegraphics[scale=1.0]{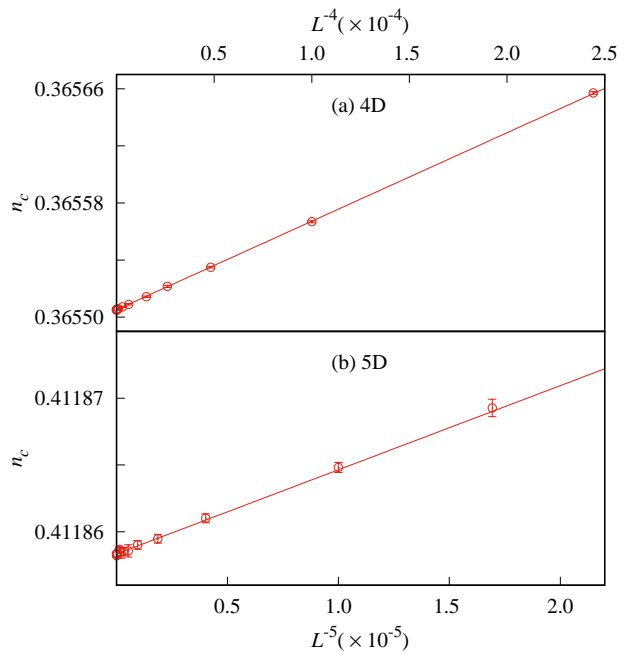}
		\caption{Plots of $n_c$ versus $L^{-4}$ or $L^{-5}$ for critical 4D (a) and 5D (b) percolation, respectively. 
		The slope of the solid lines are obtained from the fits as $0.62(1)$ and $0.62(2)$ for 4D and 5D, respectively.}
	\label{Fig:n_c}
    \end{figure}

The cluster-number density $n_c$ has its finite-size scaling as
$n_c \myeq n_{c,0}+L^{-d}\left(b+b_1L^{y_1}\right)\;$,
where $b$ is the universal excess cluster number. 
For 4D, fits with $y_1$ and $b_1$ as free fitting parameters produce estimates of 
$b_1$ being consistent with zero. 
Subsequent fits are performed with fixed $b_1 \myeq 0$.
These fitting results are summarized in Table~\ref{Tab:n_c}, 
from which we determine for 4D the excess cluster number $b \myeq 0.62(1)$ 
and the number density of critical clusters $n_{c,0} \myeq 0.365\,505\,20(3)$. 
For 5D, fits are first performed with $y_1$ being free, which lead to 
$y_1 \simeq 2.75$. 
Then fits are made with fixed $y_1=-2.75$.
The results are also listed in Table~\ref{Tab:n_c}, from which
we estimate $b \myeq 0.62(2)$ and $n_{c,0} \myeq 0.411\,858\,4(1)$. 
We plot $n_c$ versus $L^{-d}$ in Fig.\ref{Fig:n_c} for both 4D and 5D.

\begin{table}[htb]
	\centering\small
	\caption{Fitting results for the cluster-number density $n_c$.}
	\label{Tab:n_c}
	\begin{tabular}{|l|l|l|c|c|c|}
		\hline
		$d$ & $n_{c,0}$ 	& $b$ & $y_1$ 	&  $L_{\rm m}/{\rm DF}/\chi^2$ \\
		\hline
		&0.365\,505\,20(3)		&0.621(3)	 &/	&8/13/8\\
		4&0.365\,505\,20(3)	&0.617(6)   &/	&10/12/7\\
		&0.365\,505\,20(3)&0.62(1)	 &/	&12/11/7\\
		\cline{1-5}
		&0.411\,858\,44(6)		&0.611(9)  & -2.66(13)	&4/6/6\\
		5&0.411\,858\,38(7)	&0.626(13) & -3.0(3)	&5/5/5\\
		&0.411\,858\,41(4)		&0.618(2)& -2.75	&4/7/7\\
		&0.411\,858\,44(5)		&0.614(4)  & -2.75	&5/6/6\\
		\hline
	\end{tabular}
\end{table}

\subsection{Dimensionless ratios}
\begin{figure}
	\centering
	\includegraphics[scale=1.0]{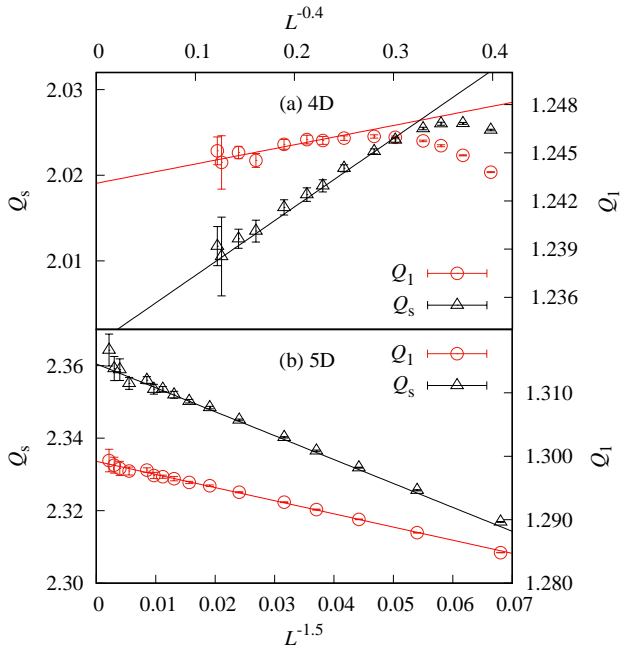}
	\caption{Plots of $Q_1$ and $Q_s$ versus $L^{-0.4}$ or $L^{-1.5}$ for critical 4D (a) and 5D (b) percolation, respectively. The straight lines are obtained from the fits.}
	\label{Fig:Q}
\end{figure}

The data of dimensionless ratios $Q_1$ and $Q_s$ are fitted to the finite-size scaling ansatz:
	\begin{equation}
		Q \myeq Q_c+ a_1L^{y_1}+a_2L^{y_2}\,\,.
	\end{equation}
In 4D, fits are first performed with $y_1$ being a free fitting parameter
and without the subleading correction term. For $Q_1$, no reasonable 
fitting results could be obtained; and for $Q_s$, $y_1$ is estimated to
be $-0.4(3)$. Then fits are performed with fixed values of $y_1$ around $-0.4$
and including the subleading correction term. In this case, reasonable results for
$Q_1$ are obtained. For $Q_1$, if including the subleading correction term with $y_2=-2$
and setting $y_1$ as a free fitting parameter, it is obtained that $y_1=-0.5(4)$.
In 5D, it is found that the data can be well described without the subleading correction
term, and fits are performed with $y_1$ being free or fixed.
Results of these fits are summarized in Table~\ref{Tab:Q}, from which  we obtain 
estimates $Q_{1,c}(4\rm D) \myeq 1.243(5)$, 
$Q_{1,c}(5\rm D) \myeq 1.299\,2(8)$, $Q_{s,c}(4\rm D) \myeq 2.00(2)$, 
and $Q_{s,c}(5\rm D) \myeq 2.360\,7(16)$. In Fig.~\ref{Fig:Q}, we plot $Q_1$ and $Q_s$ versus $L^{-0.4}$ (4D) and versus $L^{-1.5}$ (5D). 
The leading correction exponent $y_1$ for 5D is found to be much smaller than that for 4D, 
which explains why a single correction is adequate to describe the data even to 
small sizes in 5D.

\begin{table}[htb]
	
	\centering\small
	\caption{Fitting results for ratios $Q_1$ and $Q_s$.}
	\label{Tab:Q}
		\begin{tabular}{|l|l|l|c|c|c|c|}
			\hline
			$\mathcal O$ &$d$ &$\mathcal O_c$  &  $y_1$	& $y_2$  &  $L_{\rm m}/{\rm DF}/\chi^2$\\
			\hline
			&&1.242\,5(6)   & -0.4	& -1.89(14)	&8/11/4\\
			&&1.242\,9(7)	   & -0.4	& -2.1(3)	&10/10/3\\
			&4&1.242(4)	   & -0.3(2)	& -2	&8/11/4\\
			$Q_1$&&1.243(2)	   & -0.5(3)	& -2	&10/10/3\\
			\cline{2-6}
			&&1.298\,8(2)& -1.63(4) &/ &6/9/3
			\\
			&&1.298\,8(4) & -1.6(1) &/ &8/8/3
			\\
			&5&1.299\,3(1) & -1.5   &/ &8/8/5\\
			&&1.299\,1(2) & -1.5   &/ &10/7/3\\
			\hline
			&&1.996(16)   & -0.3(2)	&/	&20/7/1\\
			&&2.000(14)	& -0.4(3)	&/	&24/6/1\\
			&4&1.994\,0(17)& -0.3 & -2.6(2) &8/11/3\\
			$Q_s$ && 1.999\,8(14)& -0.4 & -2.4(2) &8/11/2\\
			\cline{2-6}
			&&2.361\,6(7)& -1.43(4) &/ &6/9/7 \\
			&&2.360\,3(12)& -1.54(10)&/ &8/8/6 \\
			&5&2.360\,8(3) & -1.5    &/ &8/8/6\\
			&&2.360\,6(5) & -1.5    & / &10/7/5\\
			\hline
	\end{tabular} 
\end{table}

\subsection{Densities of branches, junctions and non-bridges} 
\label{sec:bond-densities}
The densities are measured for three types of bonds: branches, junctions and non-bridges, 
and the data are fitted to the ansatz $\rho_0 + L^{-y_{\rho}}(a_0 + a_1 L^{y_1})$.
When setting $y_\rho$ as a free fitting parameter, fits lead to estimates  
$y_\rho ({\rm 4D}) \myeq 2.539(3)$ and $2.543(4)$ for $\rho_{\rm j}$ and $\rho_{\rm n}$, respectively; and $y_\rho ({\rm 5D}) \myeq 3.262(13)$
and $3.263(6)$ for $\rho_{\rm j}$ and $\rho_{\rm n}$, respectively. These values are consistent with 
$d-\yt \myeq 2.539(5)$, $3.263(6)$ for 4D and 5D, respectively, using our estimate of $\yt$ in Sec.~\ref{sec:the-exp}.
It is noted that the bond densities of various types are energy-density-like, 
thus the leading finite-size dependence is governed by the thermal exponent as $y_\rho \myeq d-\yt$ for all dimensions~\cite{Xu2014a,Huang2018}.
For the branch density $\rho_{\rm b}$, 
when setting $y_\rho$ as a free fitting parameter, 
no stable fits can be obtained.
And fixing $y_\rho \myeq d-\yt$, $a_0$ is found to be consistent with zero for 
both 4D and 5D. These tell that $\rho_{\rm b}$ has no significant 
finite-size dependence. 

For these bond densities, when the cutoff $L_{\rm m}$ is large, 
the parameter $a_1$ is found to be consistent with zero. 
Thus fits are performed with fixed $a_1=0$, which lead to results listed 
in Table~\ref{Tab:rho}. The results show that $a_0$ for $\rho_{\rm j}$ and $\rho_{\rm n}$ 
are equal in magnitudes within error bars and opposite in signs, for both 4D and 5D. 
This is consistent with the fact that $a_0$ for $\rho_{\rm b}$ is very close or equal to zero,
and that $\rho_{\rm b}+\rho_{\rm j}+\rho_{\rm n} \myeq p_{\rm c}$ has 
no finite-size dependence. 

\begin{table}[htb]
	
	\centering
	\caption{Fitting results for bond densities $\rho_{\rm b},\rho_{\rm j}$ and $\rho_{\rm n}$.}
	\label{Tab:rho}
	\begin{tabular}{|l|l|l|l|c|c|}
		\hline
		$d$&$\mathcal O$	& $y_\rho$ 	& $a_0$ 	&$\rho_0$ 	&  $L_{\rm m}/{\rm DF}/\chi^2$\\
		\hline
		&$\rho_{\rm b}$	&2.539		&-0.000\,05(7)			&0.147\,229\,527(19)	&12/10/5\\
		& &2.539		&0.000\,03(3)			&0.147\,229\,524(19)	&14/9/4\\
		\cline{2-6}
		&$\rho_{\rm j}$	&2.539(2)	&-0.180\,2(6)				&0.004\,637\,244(15)	&12/9/7\\
		4&		&2.539(3)	&-0.180\,1(10)				&0.004\,637\,244(16)	&14/8/7\\
		\cline{2-6}
		&$\rho_{\rm n}$&2.544(3)	&0.183\,7(14)				&0.008\,264\,449(20)	&14/8/3\\
		&		&2.543(3)	&0.182\,7(17)				&0.008\,264\,444(20)	&16/7/3\\
		\hline
		&$\rho_{\rm b}$ &3.257		&-0.000\,3(4)			&0.113\,587\,511(16)&12/7/1\\
		& &3.257		&-0.000\,3(7)			&0.113\,587\,512(16)&14/6/1\\
		\cline{2-6}
		& $\rho_{\rm j}$ 	&3.260(6)	&-0.152(2)		&0.001\,805\,416(13)	&12/6/1\\
		5&			&3.264(11)	&-0.154(4)		&0.001\,805\,415(13)	&14/5/1\\
		\cline{2-6}
		&$\rho_{\rm n}$		&3.265(3)	&0.155(1)	&0.002\,778\,517(9)		&10/7/4\\
		&			&3.262(5)	&0.155(2)	&0.002\,778\,514(10)	&12/6/3\\
		\hline
	\end{tabular}
\end{table}

\begin{figure}
 		\centering
 		\includegraphics[width=1\linewidth]{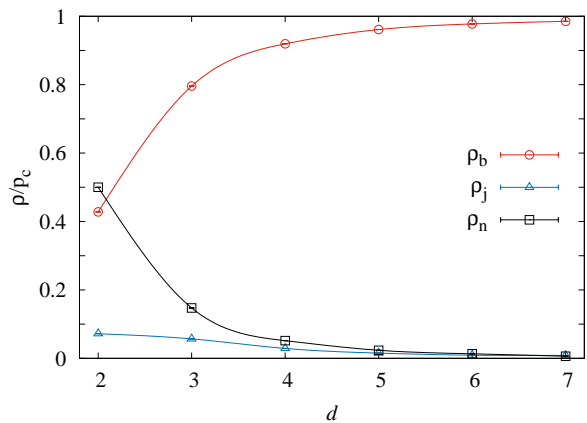}
 		\caption{Plots of the density of branches, junctions and non-bridges $\rho_{\rm b}, \rho_{\rm j}, \rho_{\rm n}$ divided by the percolation threshold $p_c$ versus the spatial dimension $d$. The lines are drawn simply to guide the eye.}
 		\label{Fig:rho_d}
 	\end{figure}

The infinite-size values of these bond densities $\rho_0$ are given in Table~\ref{Tab:d-dependent_exp},
which also includes previous results in other dimensions. 
For 3D and 6D, the results are also obtained in this work by similar methods,
but the MC simulations are less extensive than those for 4D and 5D. 
From these results we plot $d$-dependent behavior of the fraction of branches, 
junctions and non-bridges $\rho_0 / p_c$ in Fig.~\ref{Fig:rho_d}. 
The monotonically increasing (decreasing) of $\rho_{\rm b}$ ($\rho_{\rm j}$ and $\rho_{\rm n}$) 
is consistent with the common expectation
that the clusters becomes more and more dendritic as the spatial dimension $d$ increases.

 	\begin{table*}[htb]
	\centering 
	\caption{Densities of branches, junctions and non-bridges for the bond percolation on the 2D to 7D hypercubes at $p_c$. 
		The values of bond densities for 3D and 6D are also estimated in this work.}
	\label{Tab:d-dependent_exp}
	\begin{tabular}{|c|l|l|l|l|}
		\hline
		$d$ & $\rho_{\rm b}$	& $\rho_{\rm j}$	& $\rho_{\rm n}$ & $p_c$  \\
		\hline
		2 & 0.214\,050\,18(5)~\upcite{Xu2014a} & 0.035\,949\,79(8)~\upcite{Xu2014a} & 0.250\,000\,1(2)~\upcite{Xu2014a} & 1/2 \\ 
		3 & 0.198\,052(5) & 0.014\,126(6) & 0.036\,633(9) & $0.248\,811\,82(10)$~\upcite{Wang13} \\
		4 & 0.147\,229\,52(3) & 0.004\,637\,24(2) & 0.008\,264\,45(3) & $0.160\, 131\, 22(6)$~\upcite{Mertens18} \\
		5 & 0.113\,587\,51(2) & 0.001\,805\,42(2) & 0.002\,778\,51(2) & $0.118\, 171\, 45(3)$~\upcite{Mertens18} \\
		6 & 0.092\,095\,8(2) & 0.000\,863\,7(2) & 0.001\,240\,3(1) & $0.094\,201\,65(2)$~\upcite{Mertens18} \\
		7 & 0.077\,521\,1(3)~\upcite{Huang2018} & 0.000\,666\,89(3)~\upcite{Huang2018} & 0.000\,487\,24(4)~\upcite{Huang2018} & 0.078\,675\,230(2)~\upcite{Mertens18}  \\
		\hline
	\end{tabular}
\end{table*}

\section{$d$-dependence of critical exponents} 
\label{main-results}

\begin{table*}[htb]
	\centering\small
	\caption{Values of universal quantities. 
	These include estimates of four critical exponents $\yt$,$\df$,$\dB$,$\dmin$ for  percolation in $d\geq 2$. 
	Values of the excess cluster number $b$, of two dimensionless ratios $Q_1$ and $Q_s$ based on cluster-size distributions, are also included for 2D to 5D percolation. }
	\label{Tab:estiexp}
	\begin{tabular}{|c|l|l|l|l|l|l|l|l|}
		\hline
		$d$	& $y_{\rm t}$ 		& $d_{\rm f}$   & $\dB$ & $d_{\rm min}$ & $b$ & $Q_1$ &$Q_s$\\
		\hline
		2   & 3/4    & 91/48    &  1.643\,36(10)\upcite{Xu2014a} & 1.130\,77(2)\upcite{PhysRevE.86.061101} & 0.883\,576\,308\upcite{Kleban1998} & 1.041\,48(1)\upcite{Hu12} & 1.148\,69(3)\upcite{Hu12} \\
		\hline
		3   & 1.141\,30(16)\upcite{Xu2014} & 2.522\,93(10)\upcite{Xu2014} & 1.855(15)\upcite{Rintoul_1994} & 1.375\,5(3)\upcite{Xu2014} & 0.675(2)\upcite{Wang13} & 1.155\,5(3)\upcite{Wang13} & 1.578\,5(5)\upcite{Wang13} \\
		\hline 
		4  &1.459(6)\upcite{Koza16}		&3.043\,7(11)\upcite{Mertens18}		&1.95(5)\upcite{Mouk98}	&1.607(5)\upcite{Paul01} &	& & \\
		Present	&1.461(5)		&3.044\,6(7)	&1.984\,4(11)	&1.604\,2(5) & 0.62(1)	& 1.243(5) & 2.00(2) \\
		\hline
	    5 &1.747(5)\upcite{Koza16}		&3.524(2)\upcite{Mertens18}		&2.00(5)\upcite{Mouk98}	&1.812(6)\upcite{Paul01} & & &  \\
		Present &1.743(10)		&3.526\,0(14)	 &2.022\,6(27)	&1.813\,7(16)	& 0.62(2) & 1.299\,2(8)  & 2.360\,7(16) \\
		\hline
		$\geq 6$ & $d/3$\upcite{aharony1984scaling}   & $2d/3$\upcite{aharony1984scaling}    & $d/3$\upcite{aharony1984scaling} &   $d/3$\upcite{aharony1984scaling}   & & & \\
				\hline
	\end{tabular}
\end{table*}

\begin{figure}[htb]
	\centering
	\includegraphics[width=0.5\textwidth]{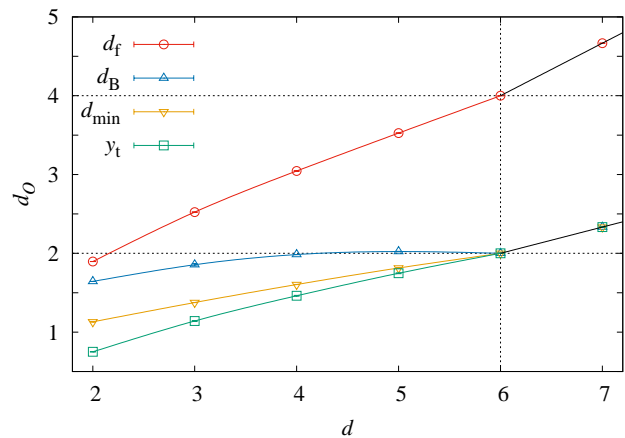}
	\caption{Plot of critical exponents $\df, \yt, \dmin, \dB$ versus the spatial dimension $d$. For $d<6$, the lines are drawn simply to guide the eye, and for $d \ge 6$ the lines are mean-field predictions.}
	\label{Fig:critexp}
\end{figure}

Table~\ref{Tab:estiexp} gives a summary of the results for the four critical exponents 
as a function of spatial dimension $d$, which is further plotted  in Fig.~\ref{Fig:critexp}.  
We obtain the following observations.  
(1), while $\yt$, $\df$ and $\dmin$ increase monotonically as $d$ becomes larger,
the backbone exponent $\dB$ exhibits a local maximum around $d \approx 5$. 
(2),  the fractal dimension of a percolation cluster $\df$ is larger than the backbone dimension $\dB$ for all $d \ge 2$, 
and as $d$ goes higher, the difference between $\df$ and $\dB$ becomes larger. 
This means that as $d$  increases, the fraction of backbones in a percolation cluster 
decreases and thus the cluster becomes more tree-like. 
This is consistent with the result that as $d$ increases, 
the fraction of the branches in the occupied bonds quickly
approaches 1; see Fig.~\ref{Fig:rho_d}.
(3), while one has $\yt < \dmin < \dB $ for $2 \le d < 6$,
these three critical exponents approach to each other as $d$ increases, 
and become identical at the upper critical dimensionality $d_{\rm u}=6$.
This means that as $d$ increases, the blobs (bridge-free clusters) become less and less compact,
and for $d \geq 6$, a blob only contains a few cycles and can break up after removing only a few occupied bonds. 
Therefore, we obtain an overall scenario that as $d$ increases, 
the structure of the percolation clusters becomes more and more tree-like, 
decorated with blobs that are less and less compact.

In addition to MC simulations, one may also utilize other methods to 
calculate the $d$-dependence of critical exponents. 
A comparison is made between the MC and $\epsilon$-expansion results
as below.

\begin{figure}
	\centering
	\includegraphics[scale=1.0]{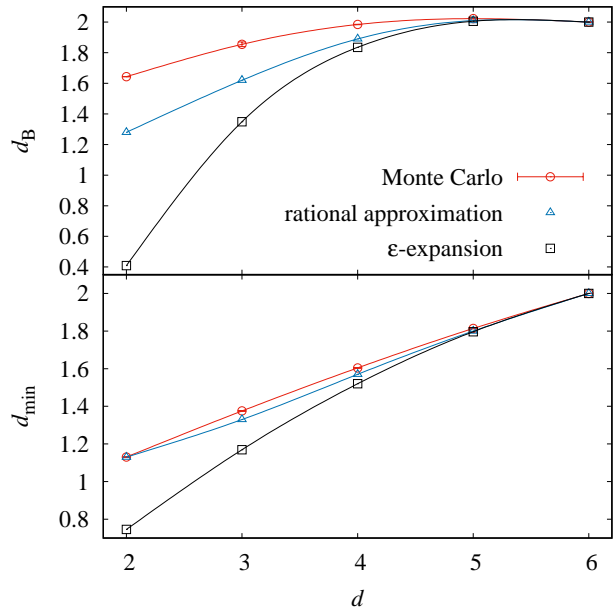}
	\caption{Dependence of $d_{\rm B}$ and $d_{\rm min}$ on the spatial dimension $d$. The $\epsilon$-expansion (squares), the rational approximation (triangles), and the numerical simulations (circles) are compared.
	The lines are drawn simply to guide the eye.}
	\label{Fig:epsilon}
\end{figure}

{\it Comparison with $\epsilon$-expansion results.}
There exists an analytic $\epsilon$-expansion method to 
approximately estimate critical exponents for dimension $d = 6 - \epsilon$
~\cite{harris1975renormalization,Gracey15, PhysRevE.59.R6239,Janssen2000}. 
For exponents $\dB$ and $\dmin$, Refs.~\cite{PhysRevE.59.R6239,Janssen2000} give the following:  
\begin{eqnarray}
	d_{\rm B} &=& 2+\frac{1}{21}\epsilon-\frac{172}{9261}\epsilon^2 \nonumber \\
	&& + 2 \frac{-74639+22680\zeta(3)}{4084101}\epsilon^3 + \bigO(\epsilon^4) \,,
	\label{Eq:dB-epsilon}
\end{eqnarray}
\begin{eqnarray}
	&&	d_{\rm min} = 2-\frac{1}{6}\epsilon \nonumber \\
	&&	-\left[\frac{937}{588}+\frac{45}{49}\left(\ln 2-\frac{9}{10}\ln 3\right)\right]\left(\frac{\epsilon}{6}\right)^2+ \bigO(\epsilon^3) \,,
	\label{Eq:dmin-epsilon}
\end{eqnarray}
where $\zeta(z)$ is the Riemann zeta function with variable $z$.
We compare the above $\epsilon$-expansion results with MC estimates 
in Fig.~\ref{Fig:epsilon}.
It is seen that the $\epsilon$-expansion becomes more and more accurate as $d$ approaches $6$. This can be expected since results from the $\epsilon$-expansion are better when $\epsilon$ is smaller. 
To reduce the discrepancy between results of the $\epsilon$-expansion
and the MC simulations in low dimensions, one can use exact results in one or
two dimensions to constrain and modify the $\epsilon$-expansion results. 
From the $\epsilon$-expansion series Eqs.~(\ref{Eq:dB-epsilon}) and (\ref{Eq:dmin-epsilon}), 
Janssen et al.~\cite{Janssen2000} make the following rational approximation for $\dB$ and $\dmin$
by incorporating the fact that they both approach one for $d \rightarrow 1$:
\begin{eqnarray}
	d_{\rm B}  \simeq 1 + \left(1-\frac{\epsilon}{5}\right)
	\left(1 + \frac{26}{105}\epsilon + \frac{7166}{231525}\epsilon^2 - 0.0170 \epsilon^3\right) \,, \nonumber \\
\end{eqnarray}
\begin{eqnarray}
	d_{\rm min} \simeq 1 + \left(1-\frac{\epsilon}{5}\right) 
	\left( 1 + \frac{\epsilon}{30} - 0.0301 \epsilon^2 \right) \,.
	\label{Eq:dmin-rational}
\end{eqnarray}
The above rational approximations for $\dB$ and $\dmin$ are also plotted in Fig.~\ref{Fig:epsilon},
from which we see that they indeed lead to results much closer to the MC simulations  
than the direct $\epsilon$-expansion Eqs.~(\ref{Eq:dB-epsilon}) and (\ref{Eq:dmin-epsilon}). 
Further, the rational approximation can be improved by resummation methods 
such as the Pad\'e and Pad\'e-Borel approximants~\cite{Bonfim81}.
For example, by making a [2,1] Pad\'e approximation,
Paul et al.~\cite{Paul01} find $\dmin=1.614$ and $1.814$ for 4D and 5D, respectively.
Comparing with $\dmin({\rm 4D})=1.568$ and $\dmin({\rm 5D})=1.803$ 
from the rational approximation Eq.~(\ref{Eq:dmin-rational}),
the Pad\'e approximation is much closer to our MC estimates: 
the 5D value $1.814$ is in good agreement with our MC result $1.813\,7(16)$,
while the 4D value $1.614$ is slightly larger than our MC estimate $1.604\,2(5)$.

For exponents $\yt$ and $\df$, estimates can be obtained from 
the $\epsilon$-expansion results of two exponents 
$\eta$ and $\eta_{\mathcal O}$ given in Ref.~\cite{Gracey15}, 
using the scaling relations $\yt = 1/\nu = 2 - \eta + \eta_{\mathcal O}$
and $\df = \yh = (d-\eta+2)/2$.
Gracey~\cite{Gracey15} makes use of exact results in one and two dimensions 
to constrain the Pad\'e approximants of the four-loop $\epsilon$-expansion series 
of exponents $\eta$ and $\eta_{\mathcal O}$.
Using the scaling relations,
the constrained Pad\'e approximants lead to $\yt=1.115\,1$ (3D), $1.445\,1$ (4D), $1.740\,3$ (5D),
and $\df=2.523\,5$ (3D), $3.047\,7$ (4D), $3.528\,3$ (5D).
Comparing these results with the MC results
in Table~\ref{Tab:estiexp},  it can be seen that the discrepancy is 
in the third decimal place for $\df$ and in the second decimal place for $\yt$. 

\section{Discussion}                    
\label{Sec:dis}

We study critical bond percolation on periodic 4D and 5D hypercubes, 
and determine a set of four critical exponents, including the thermal exponent $\yt$, 
the fractal dimension $\df$, the backbone exponent $\dB$, and the shortest-path exponent $\dmin$. 
The reliability of the results is checked by explicitly plotting the finally quoted values and 
the effects of the 3-sigma deviations. 
Our results for $\yt$ and $\df$ are compatible with or slightly more precise than
the recent estimates~\cite{Koza16,Mertens18},
and the results for $\dB$ and $\dmin$ are one or two orders more precise than those in Refs.~\cite{Mouk98,Paul01}.  
It is interesting to observe that $\dB$ is not a monotonic function of $d$ and there is a local maximum near $d\approx 5$. 
The $d$-dependent behaviors of $\yt$, $\dB$ and $\dmin$, together with the densities for bridges and non-bridges, 
confirm that as $d$ increases, the percolation clusters become more and more dendritic. 

The universal values of the excess cluster number $b$ are also determined for 4D and 5D, 
from the amplitude of the correction term with $1/L^d$ in the cluster-number density; 
see Table~\ref{Tab:estiexp}. It seems that these results have not been reported yet.
Ziff et al.~\cite{ziff1999} shows that in 2D,  
the meaning of $b$ is $N_s - c \,$, where $N_s = \bigO(1)$ is the number of spanning clusters 
and $c$ is a universal quantity related to the cluster-size distribution. 
It is expected that this scenario holds true as long as $d < d_{\rm u} = 6$. 
However, it is known that for $d > 6$, the number of spanning clusters at percolation threshold 
diverges as $N_s \sim L^{d-6}$~\cite{Aizenman97}. 
As a result, the following questions arise. 
Is the currently defined excess cluster number $b$ still a universal quantity, 
and is it related to the scaling of the number of spanning clusters?
We note that on the complete graph (CG), the leading correction term in the cluster-number density 
is of form $(\ln V)/V$~\cite{Huang2018}. 
This might hint that the excess cluster number $b$ is not well defined for $d > 6$. 

The dimensionless ratios $Q_1$ and $Q_s$ reflect the cluster-size distributions for percolation.
In addition to 4D and 5D, we have also determined the critical dimensional ratios for 7D, 
which are $Q_1({\rm 7D})=1.32(7)$ and $Q_s({\rm 7D})=2.6(1)$. 
In Ref.~\cite{Huang2018}, it is shown that at  the percolation threshold, the finite-size probability distributions of cluster sizes 
for 7D and for the CG obey the same scaling function apart from a non-universal factor. 
This suggests that as the critical exponents, the values of $Q_1$ and $Q_s$ should be identical to those 
for the CG percolation. 
Indeed, our simulations on the CG give $Q_1 = 1.327\, 0 (2)$ and $Q_s=2.622 \, 0(5)$, 
in excellent agreement with the 7D results.

\section{Acknowledgments}
H. H. acknowledges the support by the National Science Foundation of China under Grant No.~11905001 and by the Anhui Provincial Natural Science Foundation of China under Grant No.~1908085QA23. 
Y. D. acknowledges the support by the National Key R\&D Program of China under Grant No.~2016YFA0301604 and by the National Natural Science Foundation of China under Grant No.~11625522. 
We would like to thank R. M. Ziff for his helpful comments.


\begin{thebibliography}{9}
\bibitem{Hu12} H. Hu, H. W. J. Bl\"ote, Y. J. Deng, {\it Percolation in the canonical ensemble}, J. Phys. A: Math. Theor. {\bf 45}, 494006 (2012). (We note that the data presented in this JPA paper is actually for the inverse of $Q_1$, i.e. in the paper the definition should be corrected as $S_1^2/{\left<\mathcal C_1^2\right>}$.)
\end{thebibliography}
\end{document}